\documentclass[aps,prb,twocolumn, amsmath,amssymb, superscriptaddress, biblatex]{revtex4-2}
\usepackage[english]{babel}

\usepackage{graphicx}
\usepackage{dcolumn}
\usepackage{bm}
\usepackage[colorlinks=true, breaklinks=true]{hyperref}
\usepackage{physics}

\begin{document}


\title{Induced charge generated by a Coulomb impurity in transition metal dichalcogenides}

\author{V.\,K.~Ivanov}
    \email[E-mail: ]{vladislav.ivanov@metalab.ifmo.ru}
    \affiliation{School of Physics and Engineering, ITMO University, 
    197101 St.\,Petersburg, Russia}

\author{I.\,S.~Terekhov}
    \email[E-mail: ]{i.s.terekhov@gmail.com}
    \affiliation{School of Physics and Engineering, ITMO University, 
    197101 St.\,Petersburg, Russia}

\date{\today}

\begin{abstract}
We calculate analytically the charge density $\rho_{\text{ind}}(\bm r)$ at small distances induced by a Coulomb impurity in two-dimensional transition metal dichalcogenides. The calculations were performed exactly in the coupling constant. We found the first three leading terms of the charge density asymptotics at small distances. Using this result, we demonstrate that the impurity with charge number $Z=1$ remains subcritical at any value of the coupling constant, whereas impurities with higher charge number could become supercritical at certain coupling constant values. Such a behavior is similar to that in graphene.

\end{abstract}

\maketitle


\section{\label{sec:level1}Introduction}

In the pioneering work of Geim and Novoselov \cite{Novoselov2004} the properties of the first fabricated two-dimensional Dirac material (graphene) were investigated. This work became an incentive for a theoretical and experimental study of both graphene and other two-dimensional Dirac materials. In particular, such materials include transition metal dichalcogenides (TMD) (see, for example \cite{Xiao2012}), which, in contrast to graphene, are two-dimensional semiconductors with characteristic band gap of order of $1$ eV. The application of such materials is promising in many areas of technology, from optics and nanoelectronics, to spintronics \cite{Chernozatonskii2018}. Therefore, it is necessary to investigate the electronic properties of these materials. In particular, the response of an electron gas to the external impurities is of interest, since they affect the conductivity of the material.

In the present paper, we study the response of an electron gas in TMD to an external Coulomb impurity, namely, we calculate the induced charge density. The induced charge can be described as a polarization of the electron gas ground state by an external field. This process is analogous to the vacuum polarization, studied in quantum electrodynamics in the field of an atomic nucleus (see \cite{Wichmann1956, Brown1975, Milstein1983, Zeldovich1972}), where the occurrence of an induced charge leads to modification of the potential on the distances of the Compton wavelength of an electron. In the case of Dirac materials, the induced charge in graphene was discussed in many papers (see \cite{DiVincenzo1984, Nomura2007, Ando2006, Hwang2007, Katsnelson2006, Shytov2007, 
Pereira2007, Biswas2007, Fogler2007, Terekhov2008, Kotov2008}, see also a review \cite{Pereira2009}).
For TMD, in the works \cite{Voronina2019, Sveshnikov2019}, numerical results were obtained for the induced charge density with exact account of an impurity field. Moreover, in these papers the authors investigated the behavior of the induced charge in supercritical fields, in which the field creates an electron-hole pair.

We calculate analytically the induced charge density induced by a Coulomb impurity on distances much smaller than the Compton wavelength $\lambda$ of an electron in TMD. The calculations are performed exactly in the impurity field, using the convenient integral representation of an electron Green's function in a Coulomb field \cite{Terekhov2008} and calculation techniques developed in \cite{Milstein1983}. We find the leading, next-to-leading, and next-to-next-to-leading terms of asymptotics in the parameter $r/\lambda$ of the induced charge density. We find the value of a charge, localized on the impurity, exactly in the impurity field. This allows us to demonstrate, that when the value impurity field is in the vicinity of the critical one, the induced and impurity charges become comparable. Moreover, we show that, similar to the graphene case, for an impurity with charge number $Z=1$, its field remains subcritical at any strengths of the field.

\section{Induced charge density}

The induced charge density can be expressed using an electron Green's function in an external field:
\begin{eqnarray}\label{Inducedrho}
    \rho_{\text{ind}}({\bm r}) = -ieN \int\limits_{C} \frac{d\epsilon}{2\pi}\mathrm{Tr}\{
    G({\bm r},{\bm r}|\epsilon)\} \,,
\end{eqnarray}
\noindent where $e$ is an electron charge, while a coefficient $N=4$ reflects spin and valley degeneracy. The integration contour depends on the chemical potential value (see below). The electron Green's function $G({\bm r},{\bm r}'|\epsilon)$ in a Coulomb field for TDM obeys the next equation \cite{Xiao2012}: 
\begin{equation}\label{gfe} 
    \left[\epsilon -v_F\sigma\cdot\bm p
    -\Delta\sigma_z +\frac{Ze^2}{r}\right]G({\bm r},{\bm r}'|\epsilon)=\delta(\bm r-{\bm r}'),
\end{equation} 
\noindent where ${\bm p}=-i\hbar \left(\frac{\partial}{\partial x},\frac{\partial}{\partial y}\right)$ is a momentum operator,  ${\bm{\sigma}}=(\sigma_x,\sigma_y)$,  $\sigma_{i}$ are Pauli $\sigma$-matrices,  $v_F$ is a constant with the dimension of velocity, which depends on the material,  $2\Delta$ is the band gap,  $Z|e|$ is the impurity charge. In \eqref{gfe} we neglect the spin-orbit interaction, since the energy of this interaction is much smaller than $\Delta$, see \cite{Xiao2012}. 

In \cite{Terekhov2008} a convenient integral representation of Green's function was found, see \cite[Eq.(5)]{Terekhov2008}. Using this representation and equation \eqref{Inducedrho}, the authors obtained the following expression for the induced charge density in two-dimensional Dirac materials:
\begin{eqnarray}
    \label{InducedGhargeR}
    \rho_{\text{ind}}(r)&=&-\frac{eN\Delta}{\hbar v_F \pi^2
    r}\sum_{l=0}^{\infty}\int\limits_{0}^{\infty}\int\limits_{0}^{\infty}
     d\epsilon\, ds\,f(s,y)\,, 
\end{eqnarray} 
\noindent where
\begin{eqnarray}
    \label{f(sy)}
    f(s,y)&=&e^{-y\cosh s}\bigg( 2Z\alpha \cos(\mu s)\coth s I_{2\gamma}(y)\nonumber\\
    &-&\sin(\mu s)\frac{\epsilon y}{k} I'_{2\gamma}(y)\bigg)\, ,
\end{eqnarray} 
\noindent $k=\sqrt{\epsilon^2+1}\,$, $y=2r k\Delta/(\hbar v_F\sinh s)$,
$\mu=2Z\alpha\epsilon/k\,$, $I_{\gamma}(y)$ is the modified Bessel function of the 1-st kind, $I'_{2\gamma}(y)=dI_{2\gamma}(y)/dy$,  $\gamma =\sqrt{(l+1/2)^2-(Z\alpha)^2}$,  $\alpha=e^2/(\hbar v_F\varepsilon)$, where $\varepsilon$ is a permittivity of the substrate. When obtaining this formula, the integration contour $C$ in \eqref{Inducedrho} was chosen in a way, that it goes below the real axis in the left half-plane of the complex variable $\epsilon$,  crosses the real axis at $\epsilon=0$, and then goes above the real axis in the right half-plane. This corresponds to the valence band being fully occupied, i.e. we set the value of the chemical potential to zero. Note, that the coupling constant $\alpha$ depends on the material and substrate via $v_F$ and $\varepsilon$; moreover, $\alpha$ is not small since $v_F\sim c/200$, where $c$ is the speed of light \cite{Xiao2012}. Below we set  $\hbar=v_F=1$. 

The expression \eqref{InducedGhargeR} is not well-defined \cite{Milstein1983, Terekhov2008}, since it contains divergencies when energy is large and parameter $s$ is small.  To obtain a regular expression, it is necessary to renormalize the charge density \eqref{InducedGhargeR}\cite{Terekhov2008}. We imply, that the valence band is fully occupied, while the conductance band is empty (chemical potential is equal to zero). Therefore, for subcritical external field, the total induced charge should be equal to zero. Since there is no empty states in the valence band, the renormalization condition takes the following form:
\begin{eqnarray}\label{RenCond}
    Q_{\text{ind}}=\int d^2 r \, \rho_{\text{ren}}(\bm r)=0,
\end{eqnarray}
\noindent where $\rho_{\text{ren}}(\bm r)$ is the renormalized induced charge density. A detailed description of the renormalization procedure can be found in \cite{Terekhov2008, Milstein1983}, where the calculations were performed for the case of atomic nucleus in three-dimensional quantum electrodynamics and for graphene. Let us find the renormalized expression for the induced charge density. Following \cite{Milstein1983}, we go to the momentum representation:
\begin{eqnarray}
    \rho(\bm q)=\int d^2 r \, e^{i\bm q\bm r}\rho(\bm r).
\end{eqnarray}
\noindent Integrating over the directions of $\bm r$ and introducing a new variable $y = 2rk/\sinh s$, we find:
\begin{eqnarray}\label{indChargeQ}
    \rho(\bm q)&=&-\frac{eN}{\pi}\sum_{l=0}^\infty\int\limits_0^\infty\int\limits_0^\infty\int\limits_0^\infty ds\, dy\, d\epsilon  \frac{\sinh s}{k}\times\nonumber\\
    &&f(s,y)J_0\left(\frac{q y\sinh s}{2k}\right),
\end{eqnarray}
\noindent where $J_0(x)$ is the Bessel function of the 1-st kind. In momentum representation, the renormalization condition \eqref{RenCond} becomes:
\begin{eqnarray}
\label{renorm_momentum}
    \lim_{\bm q\to 0}\rho_{\text{ren}}(\bm q)=0,
\end{eqnarray}
\noindent meaning that to find the renormalized expression, we need to calculate the asymptotics of the density \eqref{indChargeQ} at $q\to 0$. For doing so, it is convenient to present the density $\rho(\bm q)$ \eqref{indChargeQ} as a sum of two contributions:
\begin{eqnarray}
    \rho(\bm q)=\rho^{(1)}(\bm q)+\rho^{(>1)}(\bm q),
\end{eqnarray}
\noindent where  $\rho^{(1)}(\bm q)$ includes terms of the first order in $Z\alpha$:
\begin{eqnarray}\label{indChargeQ1}
    \rho(\bm q)&=&-\frac{eN}{\pi}\sum_{l=0}^\infty\int\limits_0^\infty\int\limits_0^\infty\int\limits_0^\infty ds\, dy\, d\epsilon  \frac{\sinh s}{k}\times\nonumber\\
    &&f^{(1)}(s,y)J_0\left(\frac{q y\sinh s}{2k}\right),
\end{eqnarray}
\noindent where
\begin{eqnarray}
    f^{(1)}(s,y)&=&2Z\alpha e^{-y\cosh s}\bigg( \coth s I_{2l+1}(y)\nonumber\\
    &-&\frac{y\epsilon^2}{k^2} I'_{2l+1}(y)\bigg)\, ,
\end{eqnarray}
\noindent while $\rho^{(>1)}(\bm q)$ contains terms of order of $(Z\alpha)^3$ and higher:
\begin{eqnarray}
\label{rho>1Q}
    \rho^{(>1)}(\bm q)&=&-\frac{eN}{\pi}\sum_{l=0}^\infty\int\limits_0^\infty\int\limits_0^\infty\int\limits_0^\infty ds\, dy\, d\epsilon  \frac{\sinh s}{k}\times\nonumber\\
    &&\left(f(s,y)-f^{(1)}(s,y)\right)J_0\left(\frac{q y\sinh s}{2k}\right).
\end{eqnarray}

To calculate the renormalized term $\rho^{(1)}(\bm q)$, we sum over $l$, using $\sum_{l=0}^\infty I_{2l+1}(y)=\frac{\sinh y}{2}$ (see  \cite{Gradshteyn2014table}) and obtain:
\begin{eqnarray}\label{rho1}
    &&\rho^{(1)}(\bm q)=-\frac{eN}{\pi}\int\limits_{0}^{\Lambda}\frac{d\epsilon}{k}\int\limits_{s_0}^\infty ds\int\limits_0^\infty dy J_0\left(\frac{q y\sinh s}{2k}\right)\times\nonumber\\
    &&e^{-y \cosh s}\cosh s \sinh y\left(1-\frac{s y\epsilon^2}{k^2}\tanh s\coth y\right).
\end{eqnarray}
\noindent Here, we introduce the cut-off parameters in the integrals over variables $s$ and $\epsilon$: $\Lambda\gg 1$,  $s_0\ll 1$, to eliminate the divergences at large $\epsilon$ and small $s$. Then, the expression \eqref{rho1} does not contain divergences. Next, to perform renormalization we subtract from the integrand in \eqref{rho1} its asymptotics at $q\to0$, obtaining the regular expression, in which we can set $s_0=0$, $\Lambda=\infty$. Integration over $s$ and $y$ gives 
\begin{eqnarray}\label{rho1qRen}
    \rho^{(1)}_{\text{ren}}(\bm q)&=&\frac{eN\alpha}{\pi}\int\limits_0^\infty\frac{d\epsilon}{k\sqrt{1+\frac{q^2}{4k^2}}}\Bigg\{-\frac{2\epsilon^2}{q k}\sqrt{1+\frac{q^2}{4k^2}}+\nonumber\\
    &&\left(1+\frac{4\epsilon^2}{q^2}\right)\ln\left(\frac{q}{2k}+\sqrt{1+\frac{q^2}{4k^2}}\right)\Bigg\}.
\end{eqnarray}
\noindent One can check, that $\rho^{(1)}_{\text{ren}}(\bm q)$ obeys the renormalization condition \eqref{renorm_momentum}: $$\lim_{\bm q \to 0} \rho^{(1)}_{\text{ren}}(\bm q) = 0 \, . $$

\noindent In coordinate representation, the renormalized density $\rho^{(1)}_{\text{ren}}$ has the form: 
\begin{eqnarray}
\label{rho1ren}
    &&\rho^{(1)}_{\text{ren}}(\bm r)=\frac{e NZ\alpha\Delta^2}{2\pi^2R^2}\int\limits_0^\infty dx\,x\,J_0(x)\int\limits_0^\infty d\epsilon\bigg\{-\frac{2\epsilon^2 R}{x k^2}+\nonumber\\
    && \frac{1+4\epsilon^2 R^2/x^2}{k\sqrt{1+x^2/(4k^2 R^2)}}\ln\left(\frac{x}{2k R}+\sqrt{1+\frac{x^2}{4k^2R^2}}\right)\bigg\},
\end{eqnarray}
\noindent where we introduce a dimensionless parameter $R=r\Delta$.

Now let us consider the density $\rho^{(>1)}(\bm q)$ \eqref{rho>1Q}. In contrast to $\rho^{(1)}(\bm q)$, this contribution does not contain divergent integrals, however, finite renormalization is needed. Assuming $\bm q=0$ in expression for  $\rho^{(>1)}(\bm q)$ and integrating over $y$,  $\epsilon$ and  $s$, we find:
\begin{eqnarray}\label{greater3Q}
    &&\lim_{\bm q\to 0}\rho^{(>1)}(\bm q)=-\frac{2eN}{\pi}Q(Z\alpha) \,,\\
    &&Q(Z\alpha) = \sum_{l=0}^\infty \bigg[ \frac{Z\alpha}{2l+1} + Z\alpha \,\psi\left(l+\frac{1}{2}\right) \nonumber\\
    && -\frac{1}{2}\arctan\left(\frac{Z\alpha}{\gamma}\right) -  \text{Im}\left\{\ln\Gamma(\gamma +iZ\alpha)\right\} \bigg],
\end{eqnarray} 
\noindent where $\Gamma(x)$ is the Euler gamma-function,  $\psi(x)=\frac{d\ln \Gamma(x)}{dx}$ is the digamma function.  
Subtracting the term \eqref{greater3Q} from the expression for $\rho^{(>1)}(\bm q)$ \eqref{rho>1Q}, we find the renormalized expression:
\begin{eqnarray}\label{Rhogreat3RenInQ}
    \rho^{(>1)}_{\text{ren}}(\bm q)=\frac{2eN}{\pi}Q(Z\alpha) +\rho^{(>1)}(\bm q).
\end{eqnarray}
\noindent In coordinate representation, we have:
\begin{eqnarray}\label{IndCargeG3R}
    \rho^{(>1)}_{\text{ren}}&&(\bm r)=eN\Bigg(\frac{2 \,Q(Z\alpha)}{\pi}\delta(\bm r)-\nonumber\\
    \frac{\Delta}{\pi^2
    r}&&\sum_{l=0}^{\infty}\int\limits_{0}^{\infty}\int\limits_{0}^{\infty}
     d\epsilon\, ds\,\left(f(s,y)-f^{(1)}(s,y)\right)\Bigg)\,, 
\end{eqnarray}

\section{Asymptotic behavior at $r \ll 1/\Delta$}

Let us find the behavior of the induced charge density at $R \ll 1$. We begin with $\rho^{(1)}_{\text{ren}}(\bm r)$. It is convenient to represent it in a following way:
\begin{eqnarray}
    \rho^{(1)}_{\text{ren}}(\bm r)=\left(\rho^{(1)}_{\text{ren}}(\bm r)-\tilde{\rho}(\bm r)\right)+\tilde{\rho}(\bm r).
\end{eqnarray} 
\noindent Here
\begin{eqnarray}
    &&\tilde{\rho}(\bm r)=\frac{e NZ\alpha \Delta^2}{2\pi^2 R^2}\int\limits_0^\infty dx\,x\,J_0(x)\int\limits_0^\infty d\epsilon\bigg\{-\frac{2R}{x}+\nonumber\\
    && \frac{1+4\epsilon^2 R^2/x^2}{\epsilon\sqrt{1+x^2/(4\epsilon^2 R^2)}}\ln\left(\frac{x}{2\epsilon R}+\sqrt{1+\frac{x^2}{4\epsilon^2R^2}}\right)\bigg\},
\end{eqnarray}
\noindent is obtained by a substitution $k \to \epsilon$ in the integrand in \eqref{rho1ren}. The direct calculation leads to 
\begin{eqnarray}\label{Rho1tilde}
    \tilde{\rho}^{(1)}(\bm r) = eNZ\alpha\frac{\pi}{8}\,\delta(\bm r).
\end{eqnarray}
\noindent To obtain this result, we perform the Fourier transformation, integrate over $\epsilon$ and $x$, and finally return to coordinate space. 

To calculate the difference $\rho^{(1)}_{ren}(\bm r)-\tilde{\rho}^{(1)}(\bm r)$, we use the method described in details in \cite{Milstein1983}. We divide the integral over $x$ into two domains $(0,\mu)$ and $(\mu,\infty)$, and the integral over $\epsilon$ into domains $(0,\lambda)$ and $(\lambda,\infty)$,  where matching parameters $\mu$ and $\lambda$ satisfy the following conditions: 
\begin{eqnarray}
    1\ll\lambda\ll R^{-1},\quad \lambda R\ll\mu\ll 1\,.
\end{eqnarray}
\noindent By calculating the leading and next-to-leading terms of asymptotics in $R$ in each of the four domains, and summing these results, we obtain:
\begin{equation}\label{Rho1deltaRho}
    \rho^{(1)}_{\text{ren}}(\bm r)-\tilde{\rho}^{(1)}(\bm r)\approx \frac{eNZ\alpha  \Delta^2}{8}\left(2\ln R+2\gamma_E+1\right),
\end{equation}
\noindent where $\gamma_E\approx 0.577$ is the Euler constant.  By summing the contributions \eqref{Rho1deltaRho} and \eqref{Rho1tilde}, we finally obtain: 
\begin{equation}
    \rho^{(1)}_{\text{ren}}(\bm r)\approx \frac{eNZ\alpha }{8}\left(\pi\delta(\bm r) + \Delta^2[2\ln R+2\gamma_E+1]\right).
\end{equation}

To find the asymptotic behavior of $\rho_{\text{ren}}^{(>1)}(\bm r)$, we again apply the method referred above \cite{Milstein1983}. After similar procedure, we obtain:
\begin{eqnarray}
    \rho^{(>1)}_{\text{ren}}(\bm r)& \approx &eN\bigg(\frac{2}{\pi}\tilde{Q}(Z\alpha) \delta(\bm r) \nonumber\\
    &+& \frac{\Delta^2}{4} \bigg\{Z\alpha \left[\frac{R^{2\gamma_0-1}-1}{2\gamma_0-1}-\ln R\right] + B(Z\alpha)R^{2\gamma_0-1} \nonumber\\
    &+&C(Z\alpha)
    - Z\alpha\frac{2\gamma_E + 1}{2}\bigg\}
    \bigg)\,,
\end{eqnarray}
\noindent where
\begin{equation}
    \begin{aligned}
    &\tilde{Q}(Z\alpha) =  \sum\limits_{l=0}^\infty \text{Im} \left[ \ln\Gamma(\gamma-iZ\alpha) 
    + \frac{1}{2}\ln(\gamma - iZ\alpha)\right. \\
    & - (\gamma - iZ\alpha)\psi(\gamma - iZ\alpha) \bigg] +\frac{Z\alpha}{2l+1} - Z\alpha \,\psi' \left(l+\frac{1}{2} \right) ,
    \end{aligned}  
\end{equation}
\begin{eqnarray}
    B(Z\alpha)&=&\frac{4Z\alpha}{(1-2\gamma_0)\pi^2}\int\limits_{0}^{\infty}ds\frac{s}{\sinh s}\bigg(1-\nonumber\\
    &&\frac{\sqrt{\pi}\Gamma(3/2-\gamma_0)(Z\alpha s)^{\gamma_0}J_{-\gamma_0}(2Z\alpha s)}{\gamma_0\Gamma(2\gamma_0+2)\sinh^{2\gamma_0-1}s}\bigg)\,,
\end{eqnarray}
\begin{eqnarray}
    C(Z\alpha)&=& Z\alpha \frac{1}{2\gamma_0-1} +\nonumber\\
    &&\sum_{l=0}^\infty\frac{8Z\alpha}{(1-4\gamma^2)\pi^2}\int\limits_0^\infty ds\bigg(e^{-2\gamma s}\bigg[\left(1+\frac{\coth s}{2\gamma}\right)\times\nonumber\\
    &&\coth s(\cos (2Z\alpha s)+2Z\alpha s\sin (2Z\alpha s))+\nonumber\\
    &+&s\cos (2Z\alpha s)(1+2\gamma\coth s)\bigg]-\frac{1}{2\gamma s^2}\bigg)\,,
\end{eqnarray}
\noindent $\gamma_0=\sqrt{1/4-(Z\alpha)^2}$.  Note, that the coefficient $\tilde{Q}$ in front of $\delta$-function differs from $Q(Z\alpha)$ in \eqref{IndCargeG3R}, because the second term in \eqref{IndCargeG3R} also contain the $\delta$-function contribution. To find $\tilde{Q}$, it is necessary to calculate the asymptotics at $q\to\infty$ of the density \eqref{Rhogreat3RenInQ}. The $\tilde{Q}$ coincides with that obtained in \cite{Terekhov2008}.  

\begin{figure}[!htb]
    \centering
    \includegraphics[width=0.9\linewidth]{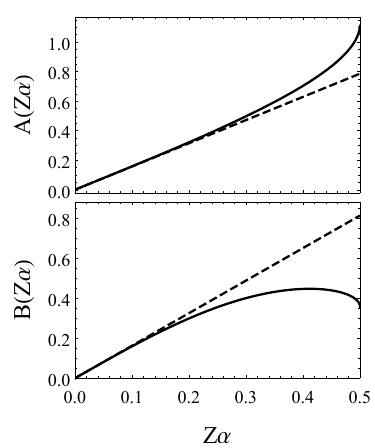}
    \caption{The functions $A(Z\alpha)$ and $B(Z\alpha)$. The solid and dashed lines correspond to exact results and linear in $Z\alpha$ results for $A(Z\alpha)$ and $B(Z\alpha)$, respectively.
    }
    \label{fig:plot}
\end{figure}

To obtain the final result, we sum up $\rho^{(1)}_{\text{ren}}(\bm r)$ and $\rho^{(>1)}_{\text{ren}}(\bm r)$, set $N=4$, and find:
\begin{eqnarray}
\label{rho_ren_final}
    \rho_{\text{ren}}(\bm r)& \approx &e \bigg(A(Z\alpha) \delta(\bm r)+\Delta^2\bigg\{Z\alpha\left[\frac{R^{2\gamma_0-1}-1}{2\gamma_0-1}\right] \nonumber\\
    &+&B(Z\alpha)R^{2\gamma_0-1}+C(Z\alpha)\bigg\}
    \bigg)\,,
\end{eqnarray}
\noindent The function $A(Z\alpha)$ was found in \cite{Terekhov2008}. The functions $A(Z\alpha)$,  $B(Z\alpha)$ and $C(Z\alpha)$ are linear in $Z\alpha$ for small $Z\alpha$:
\begin{eqnarray}
    A(Z\alpha)&\approx& \frac{\pi}{2} Z\alpha ,\\
    B(Z\alpha)&\approx& \left(-\frac{5}{2} + \gamma_E + \frac{\pi^2}{2} - 2 \ln2 \right) Z\alpha,\\
    C(Z\alpha)&\approx& \left(3 - \frac{\pi^2}{2} + 2 \ln 2 \right)Z\alpha.
\end{eqnarray}  
\noindent In the vicinity of $Z\alpha=1/2$  the functions $A(Z\alpha)$ and $B(Z\alpha)$ are regular:
\begin{eqnarray}
    A(Z\alpha)&\approx& 1.121,\\
    B(Z\alpha)&\approx& 0.356,
\end{eqnarray}
\noindent while $C(Z\alpha)$ is singular: $C(Z\alpha)\sim 1/\gamma_0$. However, the exact expression for the induced charge density $$\rho_{\text{ren}}(\bm r) = \rho_{\text{ren}}^{(1)}(\bm r) + \rho_{\text{ren}}^{(>1)}(\bm r) $$ does not contain the singularity in $Z\alpha$ at $Z\alpha=1/2$, since the singularity in $C(Z\alpha)$ is canceled with the term, which have the $R^{2\gamma_0}$ asymptotics. We do not present this term in \eqref{rho_ren_final} due to its cumbersomeness, and since it is suppressed in the parameter $R$ for $Z\alpha < 1/2$. Nevertheless, one should consider this term when the impurity charge is getting close to the critical value.  
The functions $A(Z\alpha)$ and $B(Z\alpha)$ are shown in Fig. \ref{fig:plot}. The exact results in $Z\alpha$ are shown by solid lines, and linear terms in $(Z\alpha)$ are shown by dashed lines. One can see, that $A(Z\alpha)$ becomes greater than unity when $Z\alpha \approx 0.49$. This means that the sign of the total charge localized on the impurity,
\begin{eqnarray}
    Q=|e|(Z-A(Z\alpha)),
\end{eqnarray}
\noindent  becomes negative when $Z\alpha>0.49 $ for impurities with $Z\leq 1$. Hence, for $Z = 1$ the value of the induced charge becomes large enough, so the field of the impurity never becomes supercritical. Moreover, for such charge number one cannot assume the induced charge as a response of the system to the impurity, but should solve a self-consistent problem for the potential, created by the impurity and the induced charge itself. Similar behavior of an induced charge for the case of a Coulomb impurity in graphene was discussed in \cite{Terekhov2008}. However, in TDMs the problem is more complicated because, unlike the graphene, apart from localized charge on impurity a distributed charge is present, which also must be taken into account in the self-consistent equation for the potential. This problem will be discussed later.

Clearly the behavior of the coefficient $B(Z\alpha)$ starts to differ noticeably from the linear approximation already at $Z\alpha \gtrsim 0.2$, while in the vicinity of $Z\alpha=1/2$, the linear approximation is completely inapplicable. Therefore, for values of $Z\alpha$ greater than 0.25 and should consider the field of the impurity exactly.

\section{Conclusion}

We have considered the behavior of the charge density induced by a Coulomb impurity in TMDs. We have calculated leading, next-to-leading and next-to-next-to-leading asymptotics in the parameter $r\Delta \ll 1$. The calculations have been performed exactly in $Z\alpha$. We have demonstrated, that, similar to the case of graphene, an impurity with charge number $Z=1$ does not create a supercritical field, for arbitrary value of coupling constant $\alpha$. The reason is the strong screening of the impurity. The value of the induced charge, localized on the impurity, becomes comparable to the charge of the impurity itself for $Z\alpha \approx 0.49$. For such fields, a self-consistent equation for the
potential of the impurity and the induced charge should be considered. For graphene, such equation was obtained and solved. However, in TMDs a distributed charge should be taken into account in the equation for self-consistent potential. This problem will be investigated in the future. 

Similar to the graphene, impurities with  $Z > 1$ can turn into supercritical regime ($Z\alpha>1/2$). Note, that the calculations were performed for a point-like charge, however, at the scale of the lattice spacing it is necessary to take into account the difference between real potential and the Coulomb model, which can significantly modify the induced charge behavior on these scales. This problem will also be considered in the future work.

\begin{acknowledgments}

The work of V.K. Ivanov was supported by the Russian
Science Foundation (Grant No. 22-12-00258; \footnote{\relax https://rscf.ru/en/project/22-12-00258/}). The work of I.S. Terekhov was financially supported by the ITMO Fellowship Program.

\end{acknowledgments}

\bibliography{main}

\end{document}